\documentclass[preprint,12pt]{elsarticle}

\usepackage{graphicx}
\usepackage{amsmath}
\usepackage{lineno}
\usepackage{hyperref}
\usepackage{color}

\journal{Radiation Physics and Chemistry}

\begin{document}

\begin{frontmatter}

\title{Some problems associated with the standardization of the light curve of type 1a supernovae}

\author[inst1]{A. P. Mahtessian}
\author[inst1]{G. S. Karapetian}
\author[inst2]{H. F. Khachatryan}
\author[inst2]{M. A. Hovhannisyan}
\author[inst2]{L. A. Mahtessian}
\author[inst3]{L. E. Byzalov}
\author[inst4]{J. M. Sarkissian}

\affiliation[inst1]{Byurakan Astrophysical Observatory (BAO) after V. Ambartsumian, NAS of the Republic of Armenia Byurakan, Aragatzotn Province, Republic of Armenia,0213}

\affiliation[inst2]{Institute of Applied Problems of Physics, NAS of the Republic of Armenia 25 Hrachya Nersissian Str., Yerevan, Republic of Armenia, 0014}
\affiliation[inst3]{University of Waterloo, Ontario, Canada 200 University Ave W, Waterloo, ON N2L 3G1}
\affiliation[inst4]{CSIRO, Space and Astronomy, Parkes Observatory, PO Box 276, Parkes NSW 2870, Australia}

\begin{abstract}
We show that the parameters used to standardize the luminosity of Type 1a supernovae in the SALT2 and SiFTO models are strongly dependent on the redshift z. Consequently, when standardized with increasing z, the average absolute magnitudes of Type 1a supernovae are artificially increased. This means that for a given apparent magnitude they are, on average, assigned larger distances than they actually are, creating the appearance of their recession with acceleration and requiring the introduction of the concept of antigravity (dark energy) to explain it.
We also show that after standardization, Type 1a supernovae cease to be standard candles.
We therefore argue that such a standardization is not suitable for measuring the distances to Type 1a supernovae, and hence the accelerating expansion of the Universe is called into question.
\end{abstract}


\end{frontmatter}


\section{Introduction}
Type 1a supernovae are considered standard candles. Standard candles are light sources that have the same brightness regardless of place and time. The Hubble diagram is commonly used to estimate the value of cosmological parameters. However, the scatter of points on the Hubble diagram is quite large. To reduce the scatter of points and more accurately estimate the cosmological parameters, it will be necessary to standardize the luminosities of type 1a supernovae.

In modern cosmology, the SALT2 model (\cite{Guy2007}) is most often used to standardize the luminosity of supernovae.

This model uses two parameters, X1 and C. X1 characterizes the shape of the light curve (describes the stretching of the light curve over time), and the parameter C describes the color of the supernova at maximum brightness. In particular, the distance estimate assumes that supernovae with the same color, shape, and galactic environment have, on average, the same intrinsic luminosity for all redshifts. Given these parameters, the standardization equation for SALT2 can be written as follows:
\begin{equation}
    \mu=B_{obs}-(M_B-\alpha{X1}+\beta{C})
\end{equation}
where $\mu=5log{D_L} (z)+25$  is the distance modulus, $B_{obs}$ corresponds to the peak value of the apparent stellar magnitude in the B band, and $\alpha$, $\beta$, $M_B$ are the parameters of the standardization equation to estimate the distance.

When standardizing the luminosity, it is assumed that the dependence of the supernova's brightness decay time on its maximum brightness, as well as the dependence of its peak color on its maximum brightness, do not depend on the age of the precursor.

\cite{Lee2022} show that there is a close relationship between the parameters $\alpha$, $\beta$ and the population age of host galaxies.

This dependence is that in younger galaxies, type 1a supernovae at maximum brightness have a weaker luminosity than supernovae in relatively old galaxies. Since the red shift on average characterizes the age, a dependence of these parameters on the red shift should be observed.

\cite{Rigault2020} studied the star formation rate in local environments of type 1a supernovae and found a strong dependence of the standardization parameters on the local star formation rate.

At the XV Joint Byurakan-Abastumani Colloquium we presented a paper (\cite{Mahtessian2023}) in which the dependence of the standardization parameters on the red-shift (z) is studied based on some spectroscopically detected samples of type 1a supernovae (\cite{Kowalski2008}, \cite{Amanullah2010}, \cite{Betoule2014}). All data are taken from the authors of these articles without any modifications. It is shown that there is a strong correlation between these parameters and z.

The results are shown below:
\begin{enumerate}
\item {For the \cite{Kowalski2008} sample:}
$$\Delta{M}=\alpha{X1}-\beta{C}=(0.181 \pm 0.07)z+(-0.265 \pm 0.044), t=2.41$$

Statistical significance of correlation <0.02.
\item {For the sample of \cite{Amanullah2010}:}
$$\Delta{M}=\alpha{X1}-\beta{C}=(0.193 \pm 0.059)z+(-0.210 \pm 0.030), t=3.51$$

Statistical significance of correlation <0.001.

\item {For the \cite{Betoule2014} sample:}
$$\Delta{M}=\alpha{X1}-\beta{C}=(0.264 \pm 0.039)z+(-0.005 \pm 0.016), t=6.71$$

Statistical significance of correlation $\ll{0.001}$.
\end{enumerate}

The statistical significance of the correlation is estimated as follows. We use the value

$$t=R_{\Delta{M},z}\sqrt{\frac{n-2}{1-{R_{\Delta{M},z}}^2}}$$

which is subject to the Student's distribution.

Here
$$R_{\Delta{M},z}=\frac{\sum(\Delta{M_i}-<\Delta{M}>)(z_i-<z>)}
{(n-1)\sigma_{\Delta{M}}\sigma_{z}}$$
is the correlation estimate between the values of $\Delta{M}$ and $z$, $\sigma_{\Delta{M}}$ and $\sigma_{z}$ are the standard deviations of the corresponding values.

$$\sigma_{\Delta{M}}=\sqrt{\frac{(\sum(\Delta{M_i}-<\Delta{M}>)^2}{n-1}}, \sigma_z=\sqrt{\frac{(\sum(z_i-<z>)^2}{n-1}}$$

A significant redshift dependence was also found for the X1 and C parameters separately.
More details on these results can be found in \cite{Mahtessian2023}.

To analyze the issue of dependence of standardization parameters on redshift, in this paper, we present a study of the Pan-STARRS sample of type 1a supernovae (\cite{Jones2018}), in which the supernovae were identified photometrically. We use his online supernova data table (\url{https://content.cld.iop.org/journals/0004-637X/857/1/51/revision1/apjaab6b1t2_mrt.txt}). The table contains data for 1169 supernovae, of which we did not find redshifts for two supernovae (psc140122, psc480427) in the table and removed them from consideration. As a result, 1167 type 1a supernovae remain.

We also consider the sample of \cite{Guy2010}(\url{https://cdsarc.cds.unistra.fr/ftp/J/A+A/523/A7/}), where in addition to the SALT2 type 1a supernova luminosity standardization method (\cite{Guy2007}), the SiFTO standardization method (\cite{Conley2008}) was also considered.\\
We will also show that standardizing supernova luminosities will violate the fundamental assumption that Type 1a supernovae are distance standards. To study this issue, we will use the JLA sample (Betoule et al. (2014) -\url{ https://vizier.cds.unistra.fr/viz-bin/VizieR-3?-source=J/A%2bA/568/A22/tablef3}
).\\
Here, too, all data are taken from the authors of the articles, without any changes.

\section{Results}
\subsection{Dependence of standardization parameters on redshift}
Figure 1a shows the dependence of the value $\Delta{M}=\alpha{X1}-\beta{C}$, introduced to standardize the luminosity of type 1a supernovae, on the redshift for the sample of \cite{Jones2018}.The statistical significance of the correlation is very high $(10^{-22})$.\\
Figure 1b shows the dependence of the value of $\alpha{X1}$, introduced to standardize the luminosities of type 1a supernovae, on the redshift for the sample of \cite{Jones2018}. The statistical significance of the correlation is $3.4 10^{-3}$\\  
Figure 1c shows the dependence of the value $\beta{C}$, introduced to standardize the luminosities of type 1a supernovae, on the redshift for the sample of \cite{Jones2018}. The statistical significance of the correlation is very high ($10^{-20}$).

Let us discuss the sample of supernovae studied in the work of \cite{Guy2010}. 

Figure 1d shows the dependence of $\Delta{M}=\alpha{X1}-\beta{C}$ on the redshift for the sample of \cite{Guy2010}. The statistical significance of the correlation is $10^{-6}$

Figure 1e shows the dependence of the value of $\Delta{M_{SiFTO}}=\alpha_{SiFTO}X_{SiFTO} -\beta_{SiFTO}C_{SiFTO}$ for the SiFTO method on the redshift, for the sample of \cite{Guy2010}. The statistical significance of the correlation is $10^{-17}$.

\begin{figure*}
\includegraphics[width=.9\linewidth]{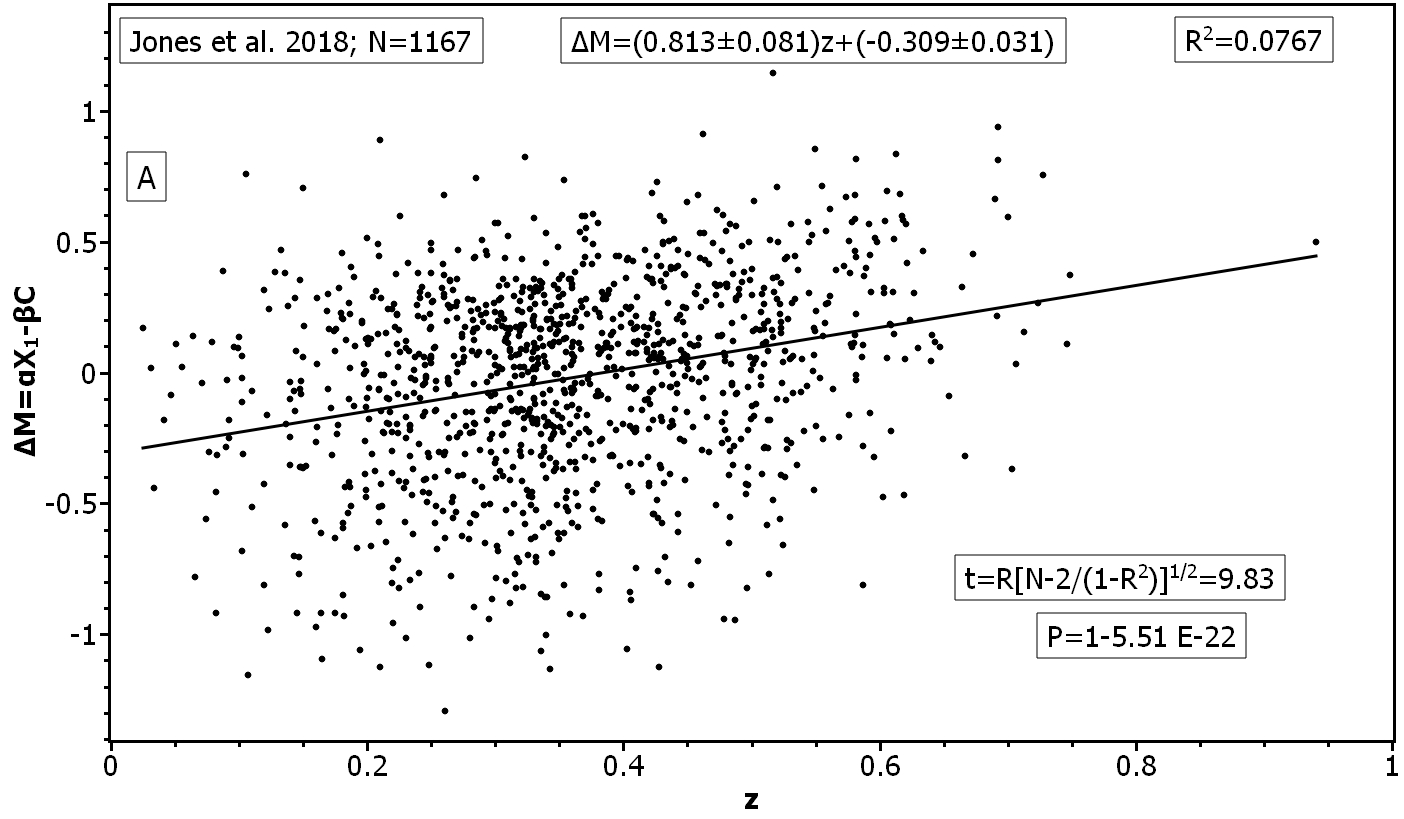} \\
 \vspace{3mm}

 \includegraphics[width=.48\linewidth]{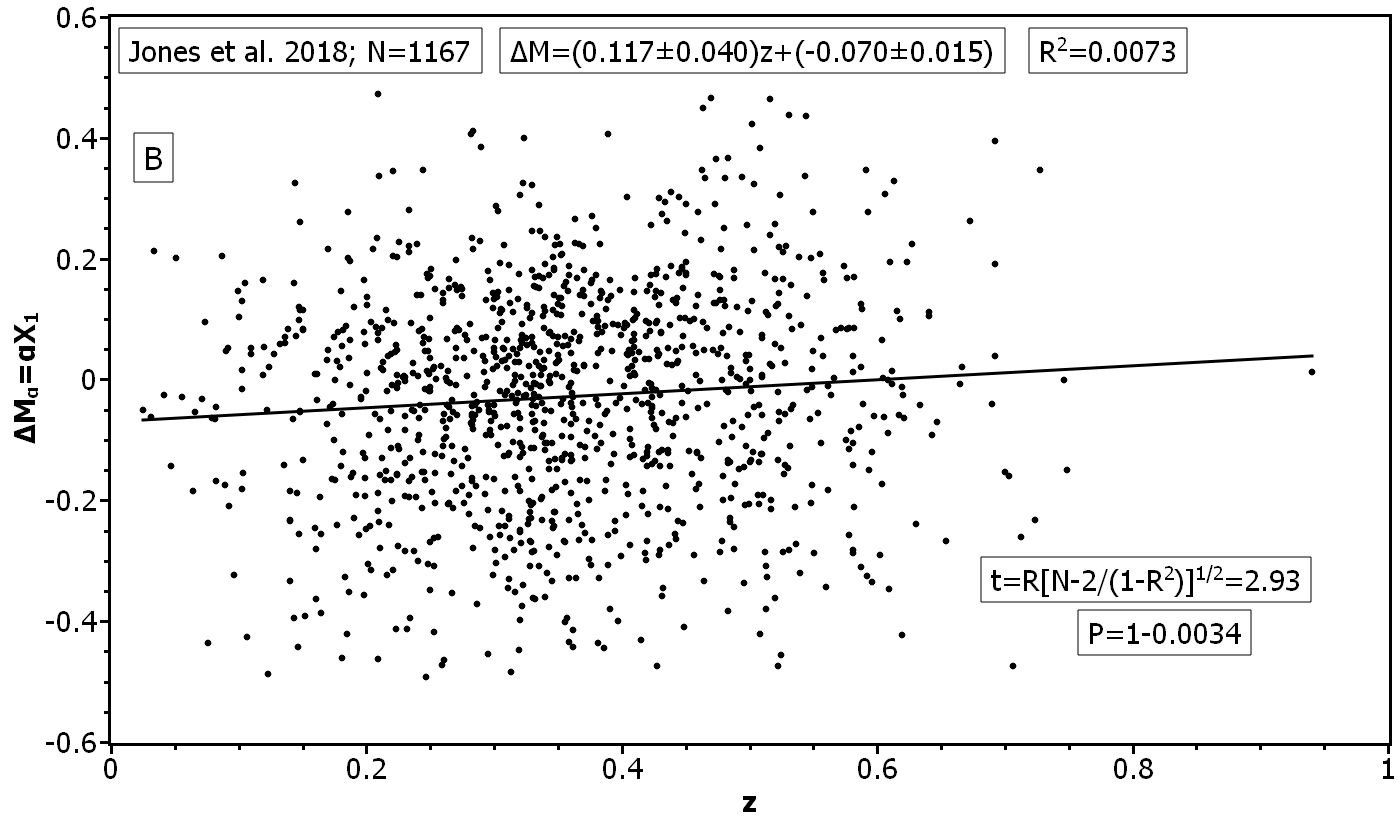} \hfill
 \includegraphics[width=.48\linewidth]{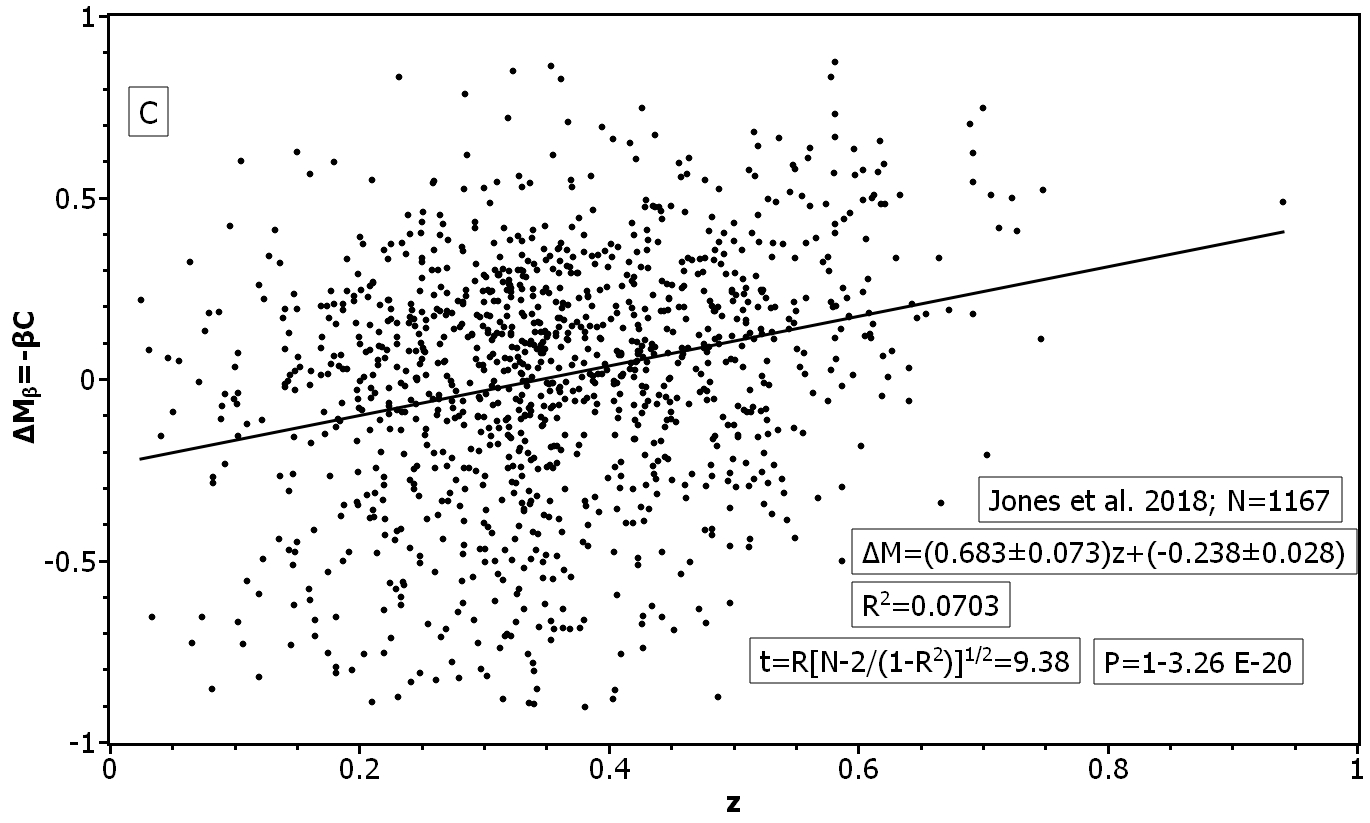} \\
 \vspace{3mm}

 \includegraphics[width=.48\linewidth]{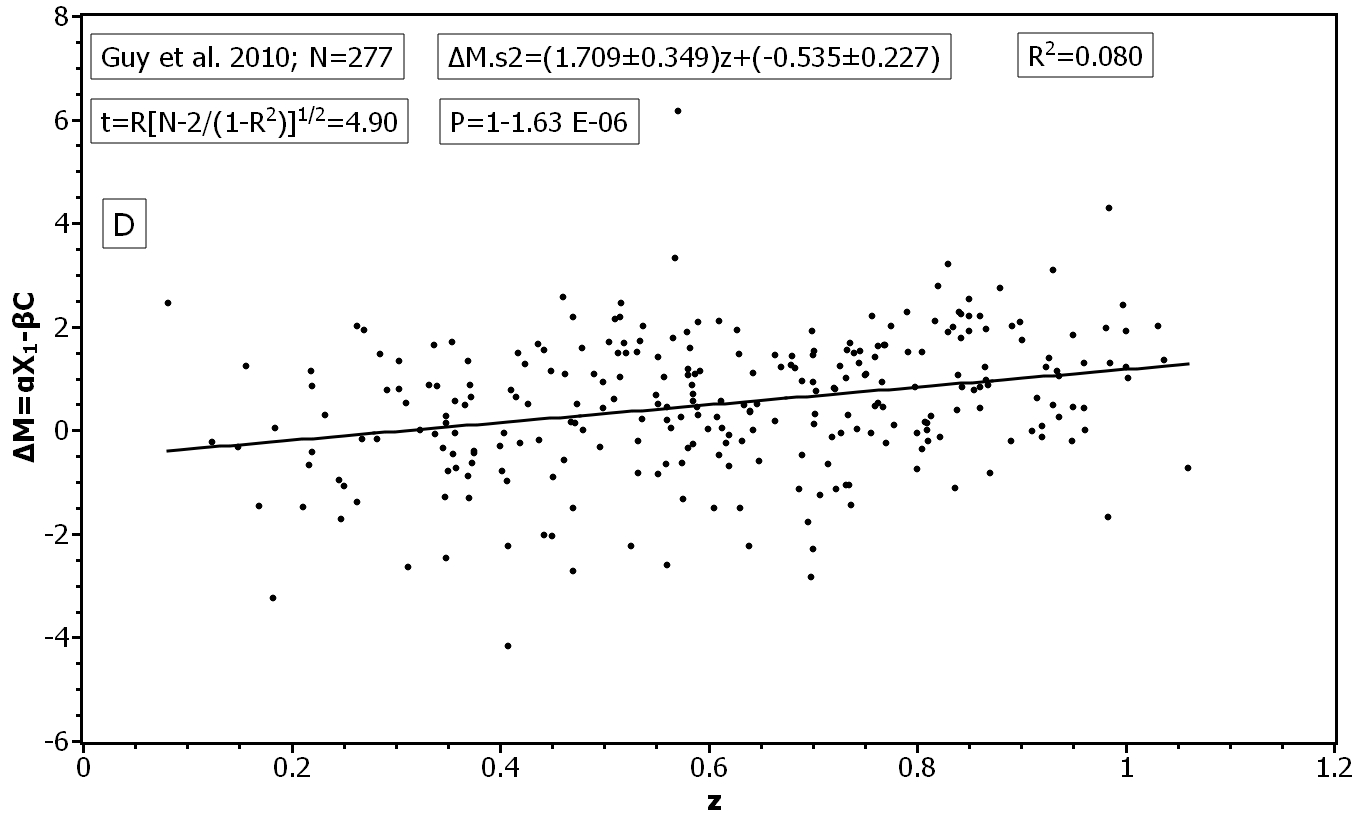}\hfill
 \includegraphics[width=.48\linewidth]{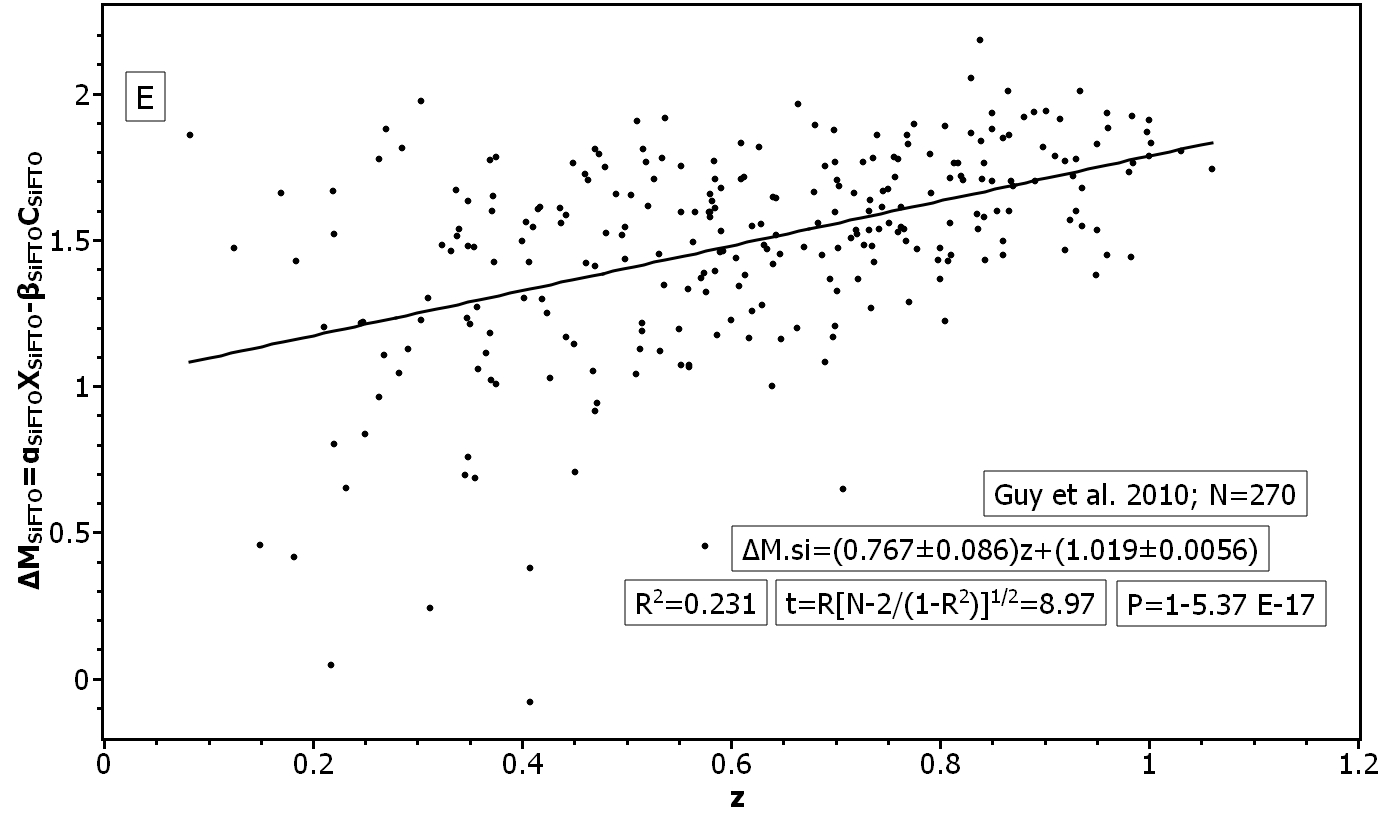}
\caption{Dependencies of the values of $\Delta{M}$, introduced to standardize the luminosities of type 1a supernovae, on the redshift.}
\end{figure*}
\subsection{Absolute magnitude test: Do Type 1a supernovae remain distance standards after standardization?}
As stated above, to reduce the scatter of points on the Hubble diagram and to estimate cosmological parameters more accurately, we should standardize the light curves of type 1a supernovae. We have seen that the standardization parameters depend strongly on the redshift. This leads us to the conclusion that such standardization is not suitable for measuring distances to type 1a supernovae.
Such a standardization may also violate a fundamental assumption about Type 1a supernovae, the standard candle assumption.
Let us see how the absolute magnitude estimates behave at different redshifts when using the obtained cosmological parameters after standardization. To study this issue, we will use, for example, the sample of \cite{Betoule2014}. 
(\url{https://vizier.cds.unistra.fr/viz-bin/VizieR-3?-source=J/A%2bA/568/A22/tablef3}).
To estimate the absolute magnitudes, the formula is used:
\begin{equation}
M=m-5lgD_L-25,                                             \end{equation} 
where $m$ is the apparent magnitude, $M$ is the absolute magnitude, $D_L$ is the Luminosity distance.
For the $\Lambda{CDM}$ model
$$
D_L(z,\Omega_M,\Omega_\Lambda,\Omega_K)=CH^{-1}_0\left(1+z\right)\left|\Omega_K\right|^{-\frac{1}{2}}\times$$
\begin{equation}
 sinn \left\{{\left|\Omega_K\right|}^{\frac{1}{2}}\int^{z}_0{dz}{\left[{\left({1+z}\right)}^{2}\left(1+\Omega_Mz\right)-z(2+z)\Omega_\Lambda \right]}^{-\frac{1}{2}}\right\}
\end{equation} 

\noindent where $z$ is the redshift of the object. $H_0$ is the Hubble constant. $\Omega_K$ is related to the curvature of space, and in the case of a flat universe it is 0 \cite{Carroll1992}: $\Omega_K=1-\Omega_M-\Omega_\Lambda$, $sinn = sinh$, when $\Omega_K \ge 0$ and $sinn = sin$, when $\Omega_K\le0$. In the case of $\Omega_K=0$, we will have:
\begin{equation}
D_L(z,\Omega_M,\Omega_\Lambda)=\frac{C(1+z)}{H_0}\int^{z}_0{dz}\left[(1+z)^2(1+\Omega_Mz)-z(2+z)\Omega_\Lambda\right]^{-\frac{1}{2}} 
\end{equation}
\cite{Betoule2014} for a flat Universe based on formula (1) and assuming that $M_0$ is independent of the redshift, obtained the following standardization coefficients: $\alpha=0.14, \beta=3.15$, and for the cosmological parameters they obtained $\Omega_\Lambda=0.705, \Omega_M=0.295$. The absolute magnitudes obtained using formulas (2) and (4) are plotted on a graph of the absolute magnitude versus redshift (Fig. 2).\\ 
It is clear from Fig. 2 that there is a strong correlation between the absolute magnitudes and the redshifts of supernovae. However, it was originally assumed that there should be no such dependence. That is, the calculation of the absolute magnitudes of supernovae using the values of cosmological parameters obtained using the luminosity standardization method significantly changed the original assumption that type 1a supernovae are distance standards.\\
\begin{figure}
	\includegraphics[width=.9\linewidth]{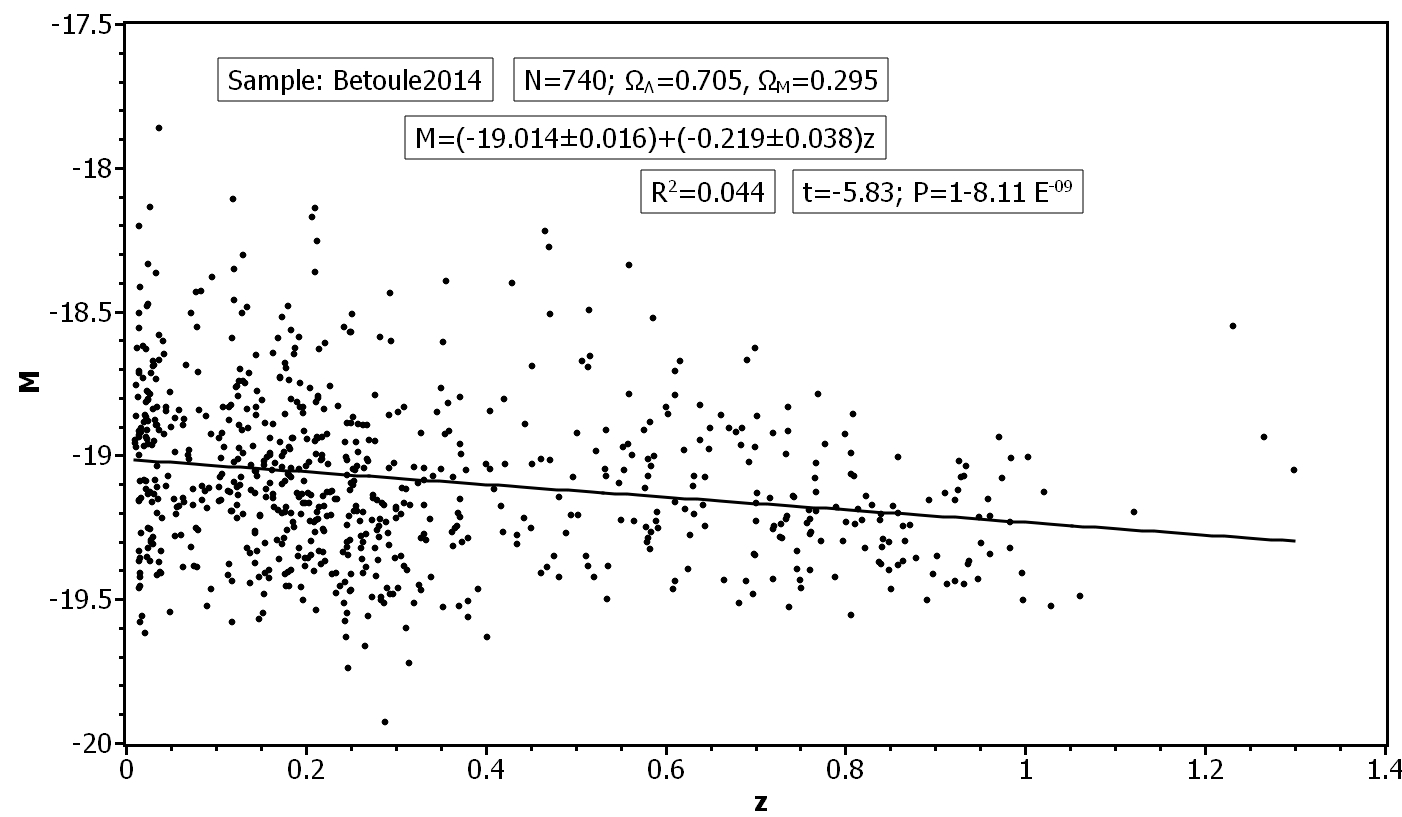}
   \caption{Dependence of the absolute magnitude ${M}$ on the redshift, for the sample of \cite{Betoule2014}, for the cosmological parameters $\Omega_\Lambda=0.705, \Omega_M=0.295$.}.
\end{figure}\\
In the work of \cite{Mahtessian2023}, the best values of the cosmological parameters for a flat universe were estimated without standardizing the luminosities and assuming that the absolute magnitudes of type 1a supernovae are independent of the redshift. There we obtained $\Omega_\Lambda=0.505, \Omega_M=0.495$. For these values of the cosmological parameters, the absolute magnitudes calculated according to (2) are plotted on the graph of the dependence of the absolute magnitude on the redshift (Fig. 3).\\
\begin{figure}
	\includegraphics[width=.9\linewidth]{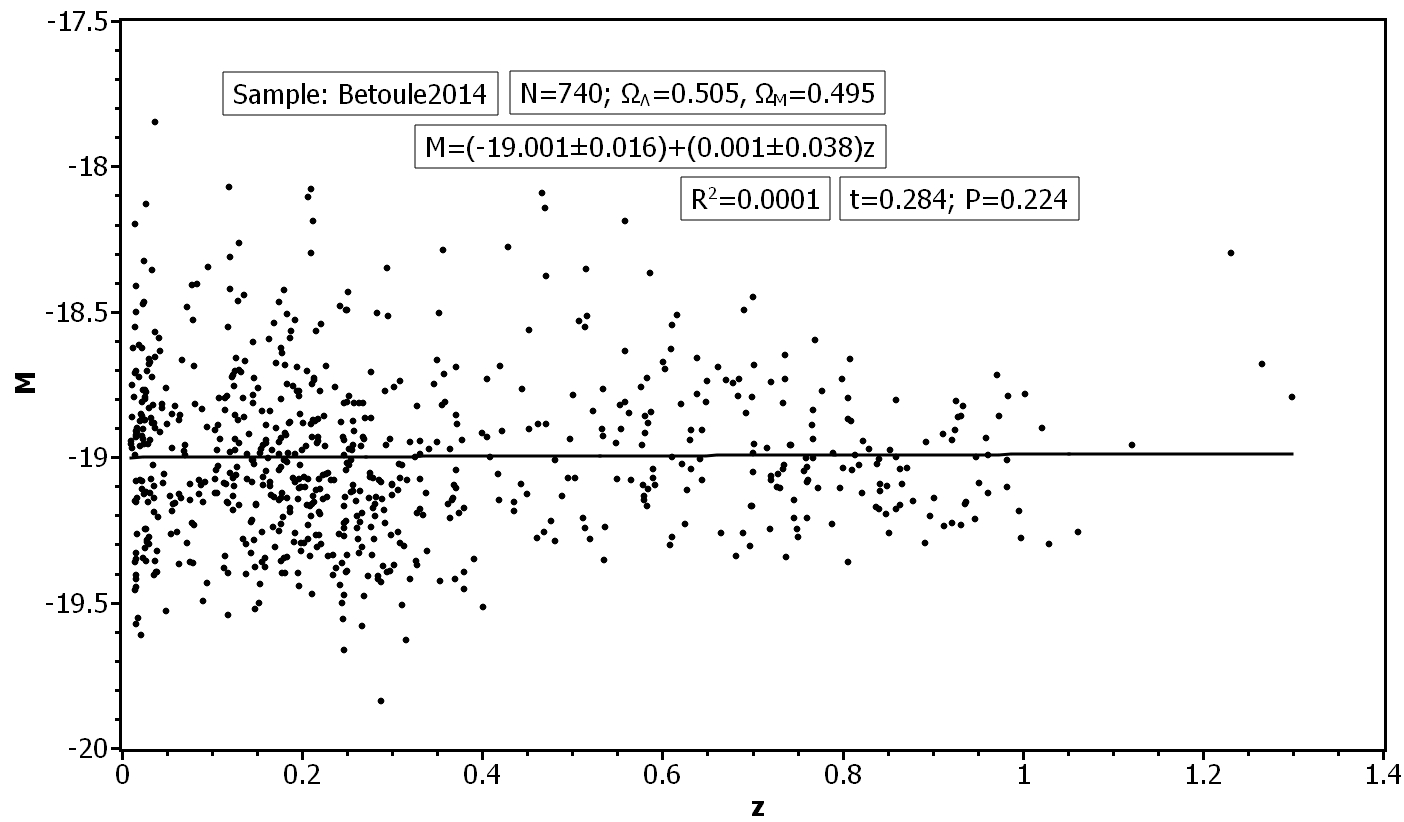}
   \caption{Dependence of the absolute magnitude ${M}$ on the redshift, for the sample of \cite{Betoule2014}, for the cosmological parameters $\Omega_\Lambda=0.505, \Omega_M=0.495$.}.
\end{figure}\\
It is evident from Fig. 3 that when using cosmological parameters estimated without luminosity standardization to estimate the distances to type 1a supernovae, the calculated absolute magnitudes of the supernovae are independent of redshift. This means that the assumption of distance standards for type 1a supernovae is not violated. It is evident that the spread of absolute magnitudes in both figures (Fig. 2 and Fig. 3) is almost the same.
Thus, standardizing supernovae also leads to a violation of the original assumption that they are distance standards. In \cite{Mahtessian2020} we tried to use the dependence of the absolute magnitude of the star with redshift to find cosmological parameters. We call this method the 'absolute magnitude test'. The idea of this test is to find cosmological parameters in such a way that the assumption that type 1a supernovae are distance standards is not violated.\\
Note that the distance standard does not necessarily have to have a constant value at all distances; it is important that a connection between its value and the distance is established. For example, one can assume that the absolute magnitude of type 1a supernovae depends on the distance, i.e. there is an evolution of the absolute magnitude. Let us estimate the values of the cosmological parameters with the assumption of the evolution of the absolute magnitude of type 1a supernovae. Then formula (1) can be written as follows:

\begin{equation}
    \mu=B_{obs}-(M_B+\epsilon{z}-\alpha{X1}+\beta{C})
\end{equation}
Here $\epsilon$ is the coefficient of dependence of the absolute magnitude on the redshift. If we do not standardize the absolute magnitudes, we can write
\begin{equation}
    \mu=B_{obs}-(M_B+\epsilon{z})
\end{equation}
We considered cases based on the sample of \cite{Betoule2014} above and try to estimate the value of cosmological parameters in the following cases:

$$A.\;\; \epsilon=0; \alpha, \beta \ne0  \;\; (Figure 2)$$ 

$$B.\;\; \epsilon=0; \alpha, \beta =0 \;\;(Figure 3)$$

$$C.\;\; \epsilon \ne0; \alpha, \beta \ne0 \;\;(Figure 4)$$

$$D.\;\; \epsilon\ne0; \alpha, \beta =0 \;\;(Figure 5)$$

We will also do an "absolute magnitude" test.
For case C, the best estimate of the cosmological parameters from the Hubble diagram fit is: $\epsilon=0.177; \alpha=0.117, \beta=2.437, \Omega_\Lambda=0.547,\Omega_M=0.453$.
For case D, the best estimate of the cosmological parameters from the Hubble diagram fit is: $\epsilon=0.361; \Omega_\Lambda=0.058,\Omega_M=0.942$ (\cite{Mahtessian2023}).
Therefore, for case C, the absolute magnitude test should show an evolution of the absolute magnitude with a slope of 0.177. In the second case, the slope should be 0.361. These cases are demonstrated in Figures 4 and 5. 

\begin{figure}
	\includegraphics[width=.9\linewidth]{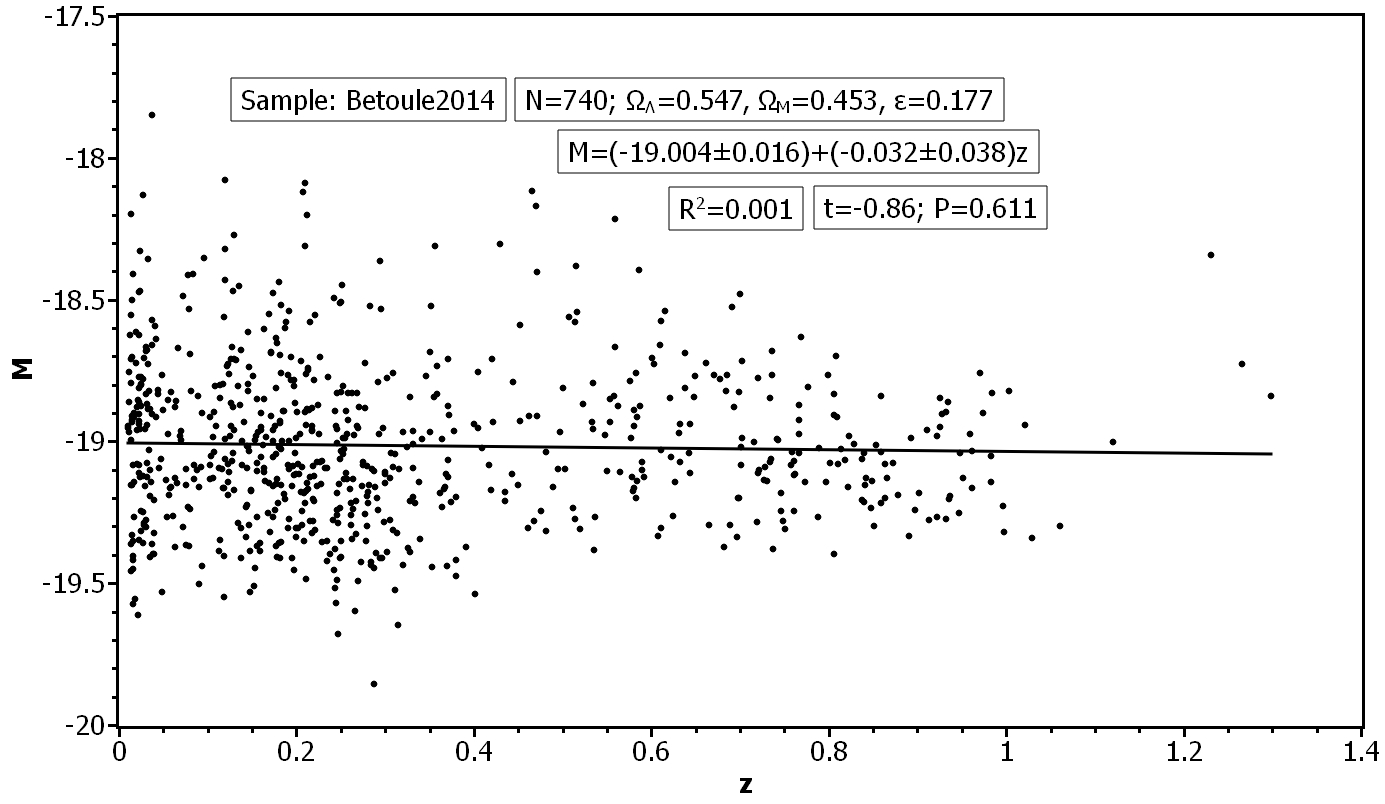}
   \caption{Dependence of the absolute magnitude ${M}$ on the redshift, for the sample of \cite{Betoule2014}, for the best fit parameters $\epsilon=0.177; \alpha=0.117, \beta=2.437, \Omega_\Lambda=0.547,\Omega_M=0.453$ (Case C).}
\end{figure}

\begin{figure}
	\includegraphics[width=.9\linewidth]{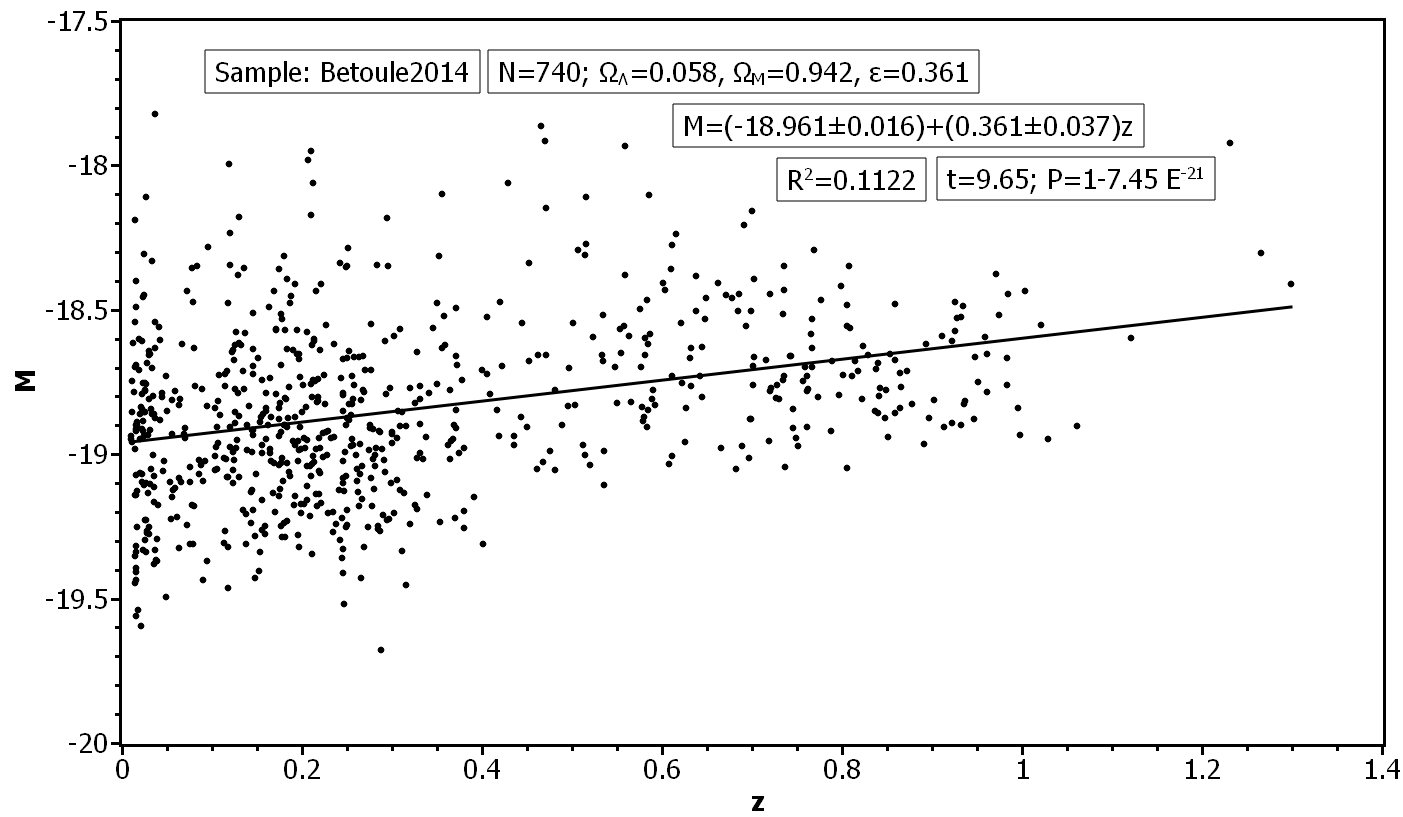}
   \caption{Dependence of the absolute magnitude ${M}$ on the redshift, for the sample of \cite{Betoule2014}, for the best fit parameters $\epsilon=0.361; \Omega_\Lambda=0.058,\Omega_M=0.942$ (Case D).}
\end{figure}

As can be seen in Figure 4 and Figure 5, in the first case, the absolute magnitude test shows a violation of the assumption of the evolution of supernova luminosities ($\epsilon \ne0.177$), and in the second case there is no violation ($\epsilon=0.361$).
Thus, standardization of luminosities violates the fundamental assumption that Type 1a supernovae are distance standards.
It is also interesting that if we assume the evolution of Type 1a supernovae luminosities without standardization of supernovae luminosities, the dark-energy density becomes almost zero. This was also noted in our other works (\cite{Mahtessian2021}, \cite{Mahtessian2023a}).

\section{Conclusion}
During this study, we obtained the following results:\\
First.\\
In all samples studied of type 1a supernovae, a strong dependence of the luminosity standardization parameters on the redshift is observed for both the SALT2 and SiFTO methods. A correlation with the redshift is also observed separately for both parameters X1 and C ($X_{SiFTO}, C_{SiFTO}$).\\
This correlation means that at large z we artificially inflate the average absolute magnitude of supernovae when standardizing, and therefore, for a given apparent magnitude, we attribute a greater distance to them on average than they actually have. So we think that they are moving away with acceleration.\\
In other words, the SALT2 and SiFTO standardization method, despite the reduction in statistical error, introduces a very large systematic error. At the same time, it is clearly seen that with efforts of the authors to reduce the statistical error, the statistical significance of the dependencies of the standardization parameters on the redshift increases.\\
Second.\\
After standardization, the obtained values of cosmological parameters violate the fundamental assumption that type 1a supernovae are distance standards. It turns out that the absolute magnitudes of supernovae calculated from these values of cosmological parameters no longer satisfy the concept of a distance standard.\\
Third.\\
If we do not standardize the luminosities, then the best value of the cosmological parameters is obtained by assuming the evolution of the supernova luminosities. But in this case the cosmological constant becomes equal to zero.\\
Thus, it can be concluded that such a standardization is not suitable for measuring distances to Type 1a supernovae.\\ 
This means that the accelerating expansion of the Universe is called into question.
\section*{Data Availability}
We used online supernova data.

\url{http://vizier.cds.unistra.fr/viz-bin/VizieR?-source=J/ApJ/686/749}

\url{https://vizier.cds.unistra.fr/viz-bin/VizieR-3?-source=J/ApJ/716/712}

\url{https://vizier.cds.unistra.fr/viz-bin/VizieR-3?-source=J/A%2bA/568/A22/tablef3}

\url{https://content.cld.iop.org/journals/0004-637X/857/1/51/revision1/apjaab6b1t2_mrt.txt}

\url{https://cdsarc.cds.unistra.fr/ftp/J/A+A/523/A7/}

\bibliographystyle{elsarticle-num}
\bibliography{references}

\end{document}